# A semi-analytical model of RF condensation that can handle localized power depositions

Ben Bobell[1,a)], Danny Sun[1], Allan H. Reiman[2]

[1]*Princeton University, Princeton, New Jersey 08544, United States of America*
[2]*Princeton Plasma Physics Laboratory, Princeton, New Jersey 08544, United States of America*
[a)] *Author to whom correspondence should be addressed: bbobell@alumni.princeton.edu*

**Abstract**

A nonlinear effect, RF (radio frequency) condensation, can be used to facilitate RF stabilization of magnetic islands. Previously studied semi-analytical models for RF condensation are suited mainly for broad deposition profiles and are unable to handle power depositions that are localized in the interior of a magnetic island. Here, a model is developed that can handle both localized profiles and broad profiles. This allows a comparison of RF condensation for narrow vs. broad deposition profiles, and it allows a study of the dependence of RF condensation of localized deposition profiles on key parameters.

## I. Introduction

Calculations in the late 1970's showing that radio frequency (RF) waves could be used to drive currents in tokamak plasmas [1, 2] were followed by calculations in the early 1980's showing that such RF driven currents could be used to stabilize magnetic islands [3, 4]. There is now an extensive theoretical literature on RF stabilization of islands along with extensive experimental demonstration of RF stabilization [5, 6, 7, 8, 9, 10, 11, 12]. Electron cyclotron current drive (ECCD) stabilization of magnetic islands is planned for ITER [13, 14, 15].

In recent years, it has been recognized that a nonlinear effect associated with the temperature perturbation in a magnetic island, called "RF condensation," can have a significant impact on the RF stabilization of magnetic islands [16]. RF condensation arises from the sensitivity of the power deposition and driven current of some RF waves to small temperature perturbations in magnetic islands. In particular, electron cyclotron waves and lower hybrid waves can display such a sensitivity because they tend to deposit their energy on the tail of the electron distribution function. The increase in the power deposition leads to a further increase in the temperature and, in turn, in the power deposition. There is a nonlinear feedback effect which leads to a nonlinearly enhanced temperature perturbation. The effect is further enhanced by the fact that the thermal diffusion coefficient in an island with flattened density and temperature profiles is much smaller than that outside the island.

RF driven current can also be sensitive to the temperature perturbation in the island, and the combined effect can lead to a concentration of the RF driven current near the center of an island, further enhancing the island stabilization effect. The condensation effect can, however, have a deleterious impact on stabilization if not properly accounted for in aiming the RF trajectory.

This paper discusses a semi-analytical model of the enhancement of the temperature perturbation arising from the nonlinear feedback on the power deposition. That nonlinear enhancement is the key to the RF condensation effect. The enhancement of the temperature perturbation can contribute not only to island stabilization via RF driven current, but also to stabilization via heating in regimes where the RF current drive is inefficient [17]

A series of papers has investigated various aspects of the physics associated with RF condensation [18, 19, 20, 21, 22, 23, 24] and a semi-analytical model has been developed that illuminates key features of the effect [16, 18, 19]. Ref. [16] presented a model that displayed a bifurcation, with the temperature increasing to infinity when the bifurcation threshold was crossed. Subsequent improvements of the model included additional physics that led to a saturation of the temperature increase above the bifurcation threshold [18, 19]. The improved model was limited in its ability to handle localized power deposition profiles within the island and was suitable primarily for broad power deposition profiles. Here we develop a model that can handle narrow deposition profiles as well as broad ones, and we use it to illuminate some key features of RF condensation for localized profiles. We apply the model to compare RF condensation for narrow vs. broad profiles, and to study the dependence of RF condensation of localized profiles on key parameters.

In Section II of this paper, we will derive our model. Section III will present solutions of the model equations and will discuss the implications of those solutions. Section IV contains a further discussion of the results and some conclusions.

## II. The Model

The RF condensation effect arises from the sensitivity of RF power deposition and current drive to the temperature perturbation in a magnetic island. When the island is heated by the RF, the increased temperature in the island leads to an increase in the power deposition in the island, and a further increase in the temperature. There is a nonlinear feedback effect that leads to a nonlinearly enhanced temperature. The sensitivity of the RF current drive in turn leads to a nonlinear increase in the driven current. In this paper we investigate a model for the nonlinear temperature enhancement.

Electron cyclotron current drive (ECCD) [25, 2, 26] and lower hybrid current drive (LHCD) [1, 27, 28, 29] are sensitive to the temperature because they deposit their energy on the tail of the electron distribution function. Considering first nonrelativistic resonant electrons, two-dimensional Fokker-Planck simulations have found that the number of such electrons is essentially determined by the lowest resonant phase velocity in the wave spectrum, both for the case of LHCD [28] and for the case of ECCD [25]. Denoting this lowest resonant phase velocity by $V_p$, the number of resonant electrons in these cases is determined by the Maxwell distribution function, and is therefore proportional to $\exp(-V_p^2/V_T^2) = \exp(-w^2)$, where $V_T$ is the thermal velocity of the electrons and we have defined $w \equiv V_p/V_T$. Writing the temperature as the sum of an unperturbed piece plus a perturbation, $T = T_0 + \tilde{T}$, the number of resonant electrons is proportional to
$n_{res} \propto \exp(-w^2) \approx \exp(-\bar{w}^2)\exp(\bar{w}^2\tilde{T}/T_0)$, where $\bar{w}$ is the unperturbed value of $w$. The RF power deposition is proportional to the number of resonant electrons,

$$P_{RF} \propto \exp(-\bar{w}^2)\exp(\bar{w}^2\tilde{T}/T_0). \qquad (1)$$

Relativistic effects need to be taken into account for ECCD [30]. For this more general case, numerical calculations find that, if we define $w_{eff}$ such that the power deposition is proportional to $\exp(-w_{eff}^2)$, then it remains the case that $w_{eff}^2 \propto 1/T$ for a broad range of parameters of interest. Thus, we can write $Tw_{eff}^2 \approx T_0\bar{w}^2$, where $\bar{w}$ is now the unperturbed value of $w_{eff}$, $\bar{w} = w_{eff}(T = T_0)$ [24]. It follows that

$$\exp(-w_{eff}^2) \approx \exp(-\bar{w}^2)\exp(\bar{w}^2\tilde{T}/T_0). \qquad (2)$$

In previous semi-analytical modeling of the condensation effect [16, 18, 19], $\bar{w}$ has been taken to be a constant, independent of position. Here we will allow $\bar{w}$ to have a spatial dependence. That will be the key to allowing localized deposition of the RF power. The spatial dependence arises, in general, through the spatial dependence of the temperature, density, and magnetic field.

We use a slab model of the island interior, corresponding to an infinitely elongated (narrow) island. We normalize lengths to the island half-width, so that the island extends from $x = -1$ to $x = 1$. $x = 0$ corresponds to the island center ("O-point"). The points at $\pm x$ correspond to the same flux surface inside the island, so that $T(x) = T(-x)$.

We adopt a simple linear model for the spatial variation of $\bar{w}$, $\bar{w}(x) = (1 - x/x_0)w_0$, where $x_0$ is the scale length of the variation of $\bar{w}(x)$ and $w_0$ is the value of $\bar{w}$ at the center of the island. The wave trajectory is assumed to traverse the island from left to right, that is, from $x = -1$ to $x = 1$. The decreasing value of $\bar{w}(x)$ along the wave trajectory is appropriate for situations where the wave is moving from a region of lower absorption to a region of greater absorption. In the case of ECCD, this corresponds to a ray trajectory that is aimed towards the electron cyclotron resonance surface.

Let $\bar{E}(x)$ be the energy density in the wave. The depletion of the energy density in the wave at any given point is proportional to the energy density remaining in the wave at that point, $\bar{E}'(x) = -\bar{\alpha}(x)\bar{E}(x)$, (we absorb the group velocity dependence into $\bar{\alpha}(x)$, so that $\bar{E}'(x)$ represents the power deposition density). The coefficient $\bar{\alpha}(x)$ is a function of the local temperature, and, following Eqs. (1) and (2) above, it can be written as $\bar{\alpha} \approx \alpha\exp(\bar{w}^2\tilde{T}/T_0)$, where $\alpha$ is a constant. ($\alpha$ does, in general have a spatial variation, but the spatial variation is dominated by that of the exponential.) We assume that the island is sufficiently large that the unperturbed temperature and density are flat in the island. We define $u$ to be a normalized temperature perturbation, $u \equiv w_0^2 \tilde{T}/T_0$. The depletion of the energy density is then given by

$$\bar{E}'(x) = -\alpha \exp(-w_0^2 x^2 / x_0^2) \exp(2w_0^2 x / x_0) \exp\left[(1-x/x_0)^2 u\right] \bar{E}(x).$$

We assume that the island grows slowly relative to the thermal confinement time in the island, so that we may use a steady state thermal diffusion equation to determine the temperature. This is justified by the fact that the local energy confinement time in a magnetic island is generally short compared to the resistive time scale on which the island grows. The diffusion equation is then $n\kappa_\perp d^2 T / dx^2 = P(x)$, where $P$ is the power density, $n$ is the density and $\kappa_\perp$ is the thermal diffusivity perpendicular to the magnetic field. It follows that $n\kappa_\perp d^2 \tilde{T}/dx^2 = 0.5\left[\bar{E}'(x) + \bar{E}'(-x)\right]$. Letting $E \equiv w_0^2 \bar{E}/(n\kappa_\perp T_0)$, we obtain our pair of model equations

$$d^2 u / dx^2 = 0.5\left[E'(x) + E'(-x)\right]. \tag{3}$$

$$E'(x) = -\alpha \exp(-w_0^2 x^2 / x_0^2) \exp(2w_0^2 x / x_0) \exp\left[(1-x/x_0)^2 u\right] E(x). \tag{4}$$

Eqs. (3) and (4) constitute the model that we are seeking. In solving the equations, the RF input energy density must be specified. We will denote this in the following by $E_1 \equiv E(-1)$. We will also take the temperature at the separatrix to be unperturbed, corresponding to $u(\pm 1) = 0$. This corresponds to the case where the RF power is initially deposited outside the island, at smaller minor radius, and is then redirected to the island.

The nonlinear effect that we are investigating enters through the $\exp\left[(1-x/x_0)^2 u\right]$ factor in Eq. (4). If this factor is absent, then the equations are linear. That is, when the factor is absent, if functions $u(x)$ and $E(x)$ provide a solution to the pair of equations, then $cu(x)$ and $cE(x)$ also provide a solution to the equations, where $c$ is a constant. In the limit that $u$ is small, Eqs. (3) and (4) reduce to the linear set of equations. This corresponds to the conventional treatment of RF heating. We will refer to this in the following as the linear limit of the equations.

The width of the deposition profile in the model is controlled primarily through $x_0$. When $x_0$ is small, the deposition tends to be localized. When $x_0$ is large the deposition profile tends to be broad. The model reduces to that of Ref. [18] in the limit here $x_0$ goes to infinity. It is possible to produce localized deposition for $x_0$ large by taking $E$ small at $x = -1$, but the deposition is then localized near $x = -1$. It is not possible to get localization in the island interior in the large $x_0$ limit.

### III. Calculations with the model

The model derived in the previous section is capable of handling deposition profiles that are localized in the island interior, unlike previous semi-analytical models. Figure 1 shows three such localized deposition profiles, each of them corresponding to an energy absorption coefficient $\alpha = 0.0002$, an inverse scale length of $x_0 = 1.2$, and a velocity ratio at the center of the island $w_0 = 3.5$. The island region extends from $x = -1$ to $x = 1$. The green curve is the deposition profile in the linear limit for an energy density $E_1 = 23$, where $E_1$ is the value of $E$ at $x = -1$. The other two curves are nonlinear deposition profiles, and they will be discussed further below.

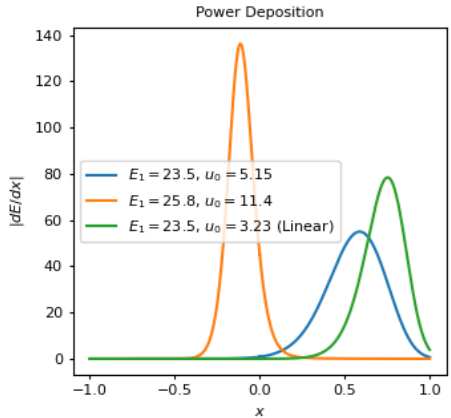

*Figure 1. Power deposition profiles in the island for $\alpha=0.0002$, $x_0 = 1.2$, and $w_0 = 3.5$. The linear calculation neglects the effect of the temperature perturbation on the power deposition. The value of the normalized temperature perturbation at the center of the island corresponding to each deposition profile is indicated in the legend.*

Figure 2 shows the normalized temperature at the center of the island as a function of $E_1$ for the values of $\alpha$, $x_0$ and $w_0$ used above. The orange line is the linear solution and the blue line is the nonlinear solution. The linear and nonlinear solutions coincide for sufficiently small $E_1$, as must be the case. As $E_1$ is increased further, the two lines begin to diverge. The increased temperature along the wave trajectory increases the power deposition there, so that the power is deposited earlier along the ray trajectory. This can be seen in the blue curve in Figure 1, which is the nonlinear deposition profile corresponding to the same energy density as the linear solution, $E_1 = 23$. The blue curve is closer to the island center, providing improved heating efficiency.

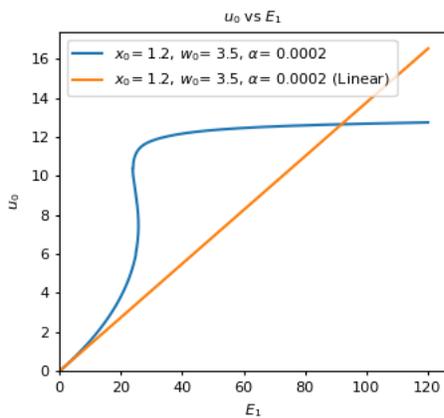

*Figure 2. Linear and nonlinear solutions for the normalized temperature at the center of the island vs. $E_1$, for $\alpha = 0.0002$, $x_0 = 1.2$, and $w_0 = 3.5$.*

As $E_1$ is increased further, a bifurcation point is encountered at a value of $E_1$ slightly above $E_1=23$. This is about twice the value of $E_1$ at which the nonlinear solution begins to deviate significantly from the linear solution. Just below the bifurcation threshold there are three branches to the solution for $u_0$. The second branch is unstable, with the temperature evolving to the value on the lower branch or top branch if the temperature is perturbed slightly below or above the value on the second branch, respectively. As $E_1$ is increased above the bifurcation threshold, there is a discontinuous jump in the steady state island temperature.

The bifurcation introduces a hysteresis effect. If $E_1$ is increased to just above the bifurcation threshold and is subsequently reduced, the solution remains on the third (top) root until a threshold in $E_1$ is encountered, at which point the temperature drops back to the first root. The threshold value of $E_1$ for dropping from the third root to the first root is lower than that for jumping from the first root to the third root. This hysteresis effect will be more pronounced in cases to be discussed below.

The blue and orange curves in Figure 1 show the nonlinear deposition profiles just below and just above the bifurcation threshold, respectively. As the bifurcation point is traversed, there is a discontinuous jump in the location of the deposition. The jump arises from a nonlinear feedback effect as the increase in the temperature earlier in the trajectory leads to increased power deposition there. The deposition profile above the bifurcation point is closer to the center of the island, so that the heating is more efficient and $u_0$ is larger.

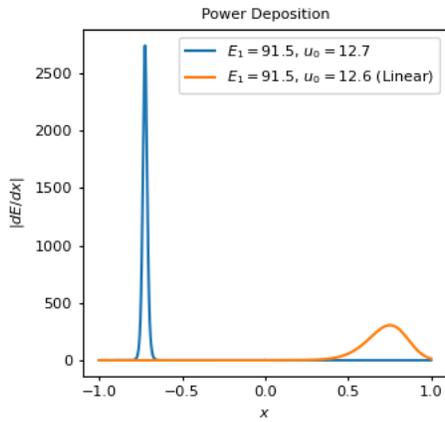

*Figure 3. Linear and nonlinear solutions for the power deposition profiles at the point where the two curves cross in Figure. 2.*

The values of $u_0$ for the linear and nonlinear solutions shown in Figure 2 cross as $E_1$ is increased further above the bifurcation threshold. To understand this, we consider the deposition profiles shown in Figure 3. The blue curve corresponds to the deposition profile for the nonlinear solution at the crossing point, and the orange curve shows the linear solution at that point. Comparing with the power deposition profiles in Figure 1, we see that the nonlinear effect causes the power to be deposited increasingly early along the ray trajectory as $E_1$ is increased, so that the heating of the island becomes increasingly inefficient, and the temperature increases only slightly despite the increase in the wave energy density, $E_1$. The linear deposition profile, on the other hand, is only changed in its amplitude, but not in its location or shape.

From a practical point of view, to deposit energy and current in an island, it will be desirable to take advantage of the narrowing of the deposition profile above the bifurcation threshold, but to avoid the decreasing effectiveness associated with further increase of the energy. In an experiment, this can be controlled by appropriate aiming of the trajectory, which is equivalent to adjusting $x_0$. For ECCD, $x_0$ roughly corresponds to the location of the resonant surface along the EC ray trajectory relative to the location of the island.

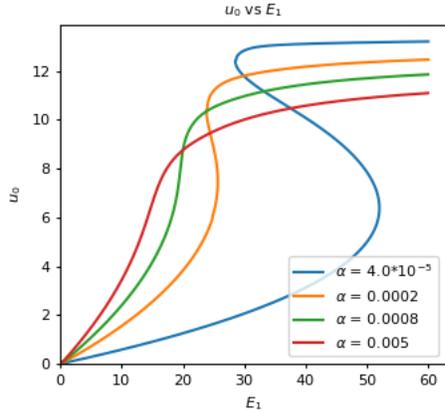

Figure 4. Dependence on α of the nonlinear $u_0$ vs. $E_1$ solutions, for $x_0 = 1.2$, and $w_0 = 3.5$.

The wave absorption coefficient in an experiment is a function of the plasma density and temperature. It is of interest to determine how the above picture is modified as the absorption coefficient changes. Figure 4 shows a set of plots of the nonlinear solutions for the normalized temperature perturbation at the center of the island as a function of $E_1$ for several different values of the absorption coefficient. We again encounter bifurcation and hysteresis. For $\alpha = 4 \times 10^{-5}$, we see that, once $E_1$ is increased above the bifurcation threshold and the third root is accessed, it is then possible to decrease $E_1$ significantly without a significant reduction in the island temperature. To see why this is so, we plot the deposition profiles on the third root for several different values of $E_1$ in Figure 5. The peak of the power deposition profile decreases in magnitude as $E_1$ is decreased, but the profile shifts closer to the center of the island, increasing the efficiency with which the temperature perturbation is produced.

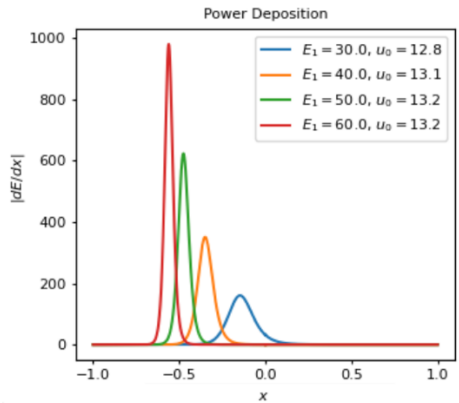

Figure 5. Power deposition profiles along the third root for $\alpha = 4 \times 10^{-5}$, $x_0 = 1.2$, and $w_0 = 3.5$.

We will see below that the value of $u_0$ at which a bifurcation is encountered in figure 4 is relatively large compared to the values at which the bifurcation is encountered for larger values of $x_0$. This dependence comes from the $(1 - x/x_0)^2$ coefficient in the exponent in Eq. 4. In this case, the linear deposition profile has $x \approx 0.8$, $1 - x/x_0 \approx 1/3$, reducing the value of the exponent by an order of magnitude.

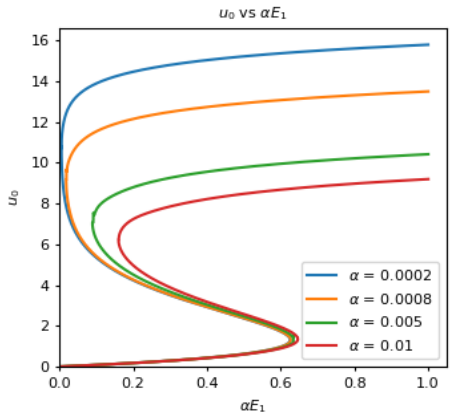

Figure 6. Dependence on α of the nonlinear $u_0$ vs. $\alpha E_1$ solutions for $x_0 = 10$ and $w_0 = 3.5$. The solutions on the first branch are approximately the same.

For large $x_0$, corresponding to relatively broad deposition profiles, it is helpful to plot $u_0$ as a function of $\alpha E_1$ rather than $E_1$. Figure 6 shows a set of such plots for several different values of α, for $x_0 = 10$. The solutions for the bottom branch are approximately the same, and the bifurcation points are approximately the same. The plots of the second and third roots diverge. This does not hold for smaller $x_0$, corresponding to more localized profiles. Figure 7 shows a plot of $u_0$ as a function of $\alpha E_1$ for $x_0 = 2$.

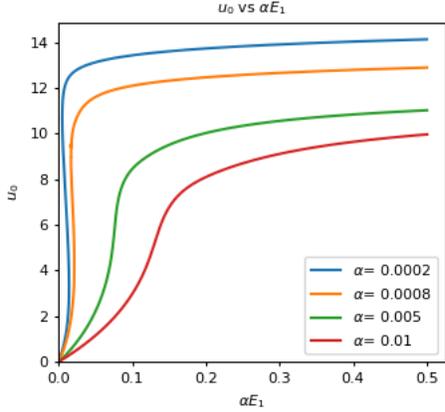

*Figure 7. Dependence on α of the nonlinear $u_0$ vs. $\alpha E_1$ solutions for $x_0 = 2$, and $w_0 = 3.5$. The solutions on the first branch are no longer approximately the same.*

The figures have thus far primarily dealt with localized deposition profiles (except for Figure 6). Figure 8 shows the effect of the transition from localized to broad profiles on the $u_0$ vs $E_1$ curves, with $x_0$ ranging from 3 to 100. For large $x_0$ (very broad profiles), the solution approaches a limit and becomes insensitive to further increases in $x_0$, with the model reducing to that studied in Ref. [18]. For large $x_0$, the value of $u_0$ at which a bifurcation is encountered asymptotes to a value of approximately 1.4.

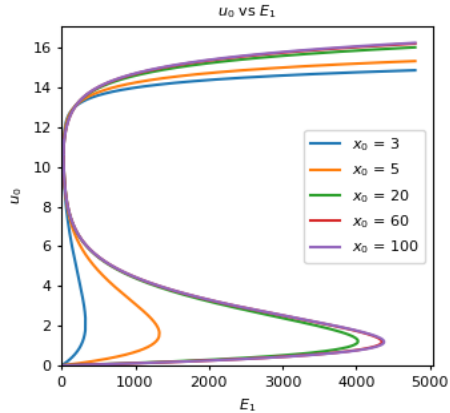

*Figure 8. Dependence on $x_0$ of the nonlinear $u_0$ vs. $E_1$ solutions for $\alpha = 4 \times 10^{-5}$ and $w_0 = 3.5$.*

The curves in figure 8 suggest that, for broad profiles, it may be desirable to take advantage of the hysteresis to stabilize islands. If $E_1$ is increased above the bifurcation threshold, it may then be decreased substantially with little decrease in $u_0$. This may aid in minimizing the impact of the island stabilization on the fusion gain

Finally, Figure 9 shows the dependence of the $u_0$ vs. $E_1$ curves on $w_0$. Although the value of $u_0$ at which a bifurcation is encountered is approximately the same for the different curves, recall that the definition of $u$ contains a factor of $w_0^2$. The magnitude of the temperature perturbation at which a bifurcation is encountered scales approximately as $1/w_0^2$.

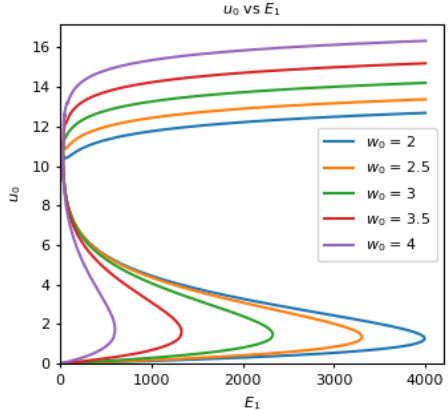

Figure 9. Dependence on $w_0$ of the nonlinear $u_0$ vs. $E_1$ solutions for $\alpha = 2 \times 10^{-4}$ and $x_0 = 5$.

## IV. Discussion

We have developed a semi-analytical model of RF condensation that can handle localized as well as broad power deposition profiles. The model has been applied to a comparison of RF condensation for narrow vs. broad profiles, and to study the dependence of RF condensation of localized profiles on key parameters. Broadness or narrowness of a deposition profile is defined in relation to the island width. In this sense, both broad and narrow deposition profiles are encountered in practice.

The model described in this paper does not directly include the effect of the stiffness of the temperature profile that is encountered when a microinstability threshold is crossed. This effect can be taken into account, however, in interpreting the predictions of the model. In the absence of heat or particle sources in an island, the diffusion equations for heat and density predict that the density and temperature profiles in a large island will be flat [31]. There is experimental [32] [33, 34, 35] and computational [36] evidence that the thermal diffusion coefficient in an island is then much smaller than the ambient thermal diffusion coefficient in the surrounding plasma. It can be expected that the diffusion coefficient will increase when the temperature gradient in the island becomes sufficiently large to encounter a microinstability threshold. With a flat density gradient, we would expect to encounter an ITG mode [37]. A bifurcation may not be seen if the $\tilde{T}/T_0$ microinstability threshold is encountered below the bifurcation threshold.

In Fig. 2, the nonlinear solution for $u_0$ begins to deviate significantly from the linear solution when $u_0$ is about half its value at the bifurcation threshold. The calculations for Figure 2 had $x_0 = 1.2$, which corresponds to localized deposition, as can be seen in Figure 1. Previous investigations for broad deposition profiles have also found a ratio of approximately 2 or 3 between the two thresholds. The nonlinear effect on the stabilization via RF driven current becomes significant at values of $u_0$ well below those at which a bifurcation is encountered. The stiffness threshold places much less of a constraint on encountering nonlinear effects than it does on encountering the bifurcation threshold. Nonlinear effects may be used to advantage even if the bifurcation threshold is not reached; and the nonlinear effects will need to be considered in aiming ray trajectories for values of $\tilde{T}/T_0$ well below those at which a bifurcation is seen.

To estimate the ITG threshold in an island, we can compare with the threshold in an axisymmetric tokamak. The threshold condition for a toroidal ITG mode in an axisymmetric tokamak is estimated as $-(R/T)(\partial T/\partial r) = \kappa_c$, where the parameter $\kappa_c$ determines the threshold [38]. Tokamak experiments suggest that $\kappa_c$ lies in the range from 3 to 8[38]. If we take this as an estimate of the ITG threshold in an island, it suggests that the ITG threshold in an island will be encountered when $\tilde{T}/T_0 \leq W/a$, where $W$ is the island width and $a$ is the minor radius of the plasma. (We have assumed a tokamak aspect ratio of 3.) In practice, very little is known about the turbulent transport in the interior of a magnetic island at this time, so that there is great uncertainty attached to this estimate. We also note that the ITG mode affects mainly the ions, so in regimes where the collisionality is sufficiently low that the ions and electrons are weakly coupled, the ITG threshold has less of an impact.

In Figure 9 it can be seen that the value of $u_0$ at the bifurcation threshold is relatively insensitive to $w_0$ for $x_0 = 5$, and that it lies between $u_0 = 1$ and $u_0 = 2$. Also in Fig. 8 the value of $u_0$ at the bifurcation threshold is relatively insensitive to $x_0$ for the range of values of $x_0$ included there. From the definition of $u_0$ it follows that the temperature perturbation at the bifurcation threshold for these parameters is approximately proportional to $1/w_0^2$. The stiffness constraint becomes less of an issue as $w_0$ is increased. This might have been surmised already from Eqs.1 and 2. An increase in the value of $w_0^2$ corresponds to the deposition of the power farther out on the tail of the electron distribution function. This is also the regime where electron cyclotron current drive (ECCD) and lower hybrid current drive (LHCD) are most efficient because the collision frequency of the accelerated electrons is lower at higher electron velocities. The economics of fusion reactors will be sensitive to the recirculating power, so that it will be desirable to drive RF-driven currents as efficiently as possible. For electron cyclotron waves, current drive is most efficient when the waves are launched from the top of the torus [39, 40]. Most contemporary tokamak experiments use outside launch, launching the waves from the low field side of the torus. Experiments on the DIII-D tokamak have demonstrated top launch efficiencies twice that of outside launch [41]. LHCD efficiencies are comparable to those of top launch ECCD. For both methods of RF current drive, values of $w^2$ close to 10 have been achieved.

The insensitivity of the threshold value of $u_0$ to $x_0$ does not continue to hold at lower values of $x_0$, corresponding to more localized deposition profiles. In Figs. 2 and 4 it can be seen that the value of $u_0$ at the bifurcation threshold for $x_0 = 1.2$ is considerably greater than at larger values of $x_0$. This effect arises from the $(1 - x/x_0)^2$ coefficient in the exponent in Eq. (4). When the stiffness threshold comes into play, it imposes an impediment to accessing the bifurcation threshold with localized deposition profiles.

In addition to the stiffness threshold, the other key parameter that comes into play in determining whether a bifurcation is seen is the magnitude of the injected wave energy density, represented by $E_1$ in our model. In Figure 8, $\alpha$ is kept fixed as $E_1$ increases, and it can be seen that increasingly large injections of power are needed to reach the bifurcation threshold as the deposition profile broadens. This is to be expected. For broad profiles, an increasing fraction of the wave power is deposited outside the island as the deposition profile broadens further. The fraction of the power deposited in the island becomes important. It can be seen from Eq. (4) that, in the linear regime, the power deposited in an island by a broad deposition profile is determined by $\alpha E_1$. Figure 6 shows solution curves for $u_0$ as $\alpha E_1$ is varied for $x_0 = 10$ and several different values of the absorption coefficient $\alpha$. The solutions as a function of the power deposited in the island, as measured by $\alpha E_1$, are almost indistinguishable until the bifurcation threshold is reached. When $\alpha E_1$ crosses the bifurcation threshold, the temperature increase along the ray trajectory causes the power to be deposited earlier in the trajectory. In particular, the power deposition in the island increases. Now the differences in $E_1$ come into play, and the solution curves diverge.

The effect of $\alpha E_1$ is different for localized deposition profiles. As can be seen in Fig. 7 for $x_0 = 2$, the solutions at the same values of $\alpha E_1$ are no longer approximately the same. For localized deposition profiles, it is $E_1$ rather than $\alpha E_1$ that determines the total power deposited in the island. In Fig. 7, the value of $E_1$ at a fixed value of $\alpha E_1$ decreases as $\alpha$ increases, and the temperature perturbation is correspondingly smaller. Even for the same values of $E_1$, differences in $\alpha$ cause the power to be deposited in different regions of the island, leading to differences in $u_0$. In Fig. 4, $u_0$ is plotted as a function of $E_1$ for $x_0 = 1.2$. In the linear regime, corresponding to sufficiently small values of $E_1$, the temperature perturbation increases as $\alpha$ increases. In the strongly nonlinear regime on the third root, the opposite is true. As the heating along the ray trajectories moves the deposition profile earlier in the trajectory, the heating becomes increasingly inefficient.

Fig. 1 shows several deposition profiles for $\alpha = 0.0002$. As $E_1$ is increased to just above the bifurcation threshold, the deposition profile jumps to a position closer to the center of the island, so that the heating becomes more efficient. The increased efficiency is reflected in the much larger value of $u_0$ in Figure 2. The power is now, however, deposited along the ray trajectory mostly before the trajectory reaches the center of the island. As $E_1$ is increased further and the deposition moves earlier in the ray trajectory, the deposition profile moves farther from the center of the island, and the heating becomes increasingly inefficient. This is also reflected in the corresponding values of $u_0$ in Figure 2.

The picture that emerges, then, is one where broader profiles are less impacted by profile stiffness above microinstability thresholds, but where localized profiles require less power to access the bifurcation threshold. Nonlinear effects begin to be significant at temperature perturbation levels well below those at which bifurcation occurs, and at lower power.

The insights provided by the model discussed in this paper provide some general guidance that will be useful when planning an experiment or a stabilization scenario. The guidance will be different, depending on whether the microinstability threshold or the power requirement is more of a constraint. For planning, we may have control over many parameters, such as the location of the wave launcher, the launching angles, the location and size of the magnetic island, the temperature and density profiles in the plasma, and the shapes of the flux surfaces. Increasingly realistic models of RF condensation are being developed that can calculate the effects of these parameter choices. The OCCAMI code couples a ray tracing code for a numerically specified tokamak equilibrium reconstruction to a solution of the nonlinear thermal diffusion equation in a magnetic island [24]. Development of the code is continuing, and it is being used for detailed experimental planning. Semi-analytic models, such as the one presented in this paper, provide insights that help to guide such planning. Planning exercises will pose questions that suggest the development of further improved semi-analytical models.

## Acknowledgments

This work was supported by DOE contracts DE-AC02-09CH11466 and DE-SC0023236.

## References

[1] N. J. Fisch, *Phys. Rev. Lett.,* vol. 41, p. 873, 1978.

[2] N. J. Fisch and A. H. Boozer, *Phys. Rev. Lett.,* vol. 45, p. 720, 1980.

[3] A. H. Reiman, *Phys. Fluids,* vol. 26, p. 1338, 1983.

[4] Y. Yoshioka, S. Kinoshita and T. Kobayashi, *Nucl. Fusion,* vol. 24, p. 565, 1984.

[5] G. Gantenbein; H. Zohm; G. Giruzzi; S. Günter; F. Leuterer; M. Maraschek,; .Meskat; Q. Yu; the ASDEX Upgrade Team, and the ECRH Group (AUG), *Phys. Rev. Lett.,* vol. 85, p. 1242, 2000.

[6] H. Zohm, G. Gantenbein, A. Gude, S. Günter, F. Leuterer, M. Maraschek, J.P. Meskat, W. Suttrop, Q. Yu, ASDEX Upgrade Team and ECRH Group (AUG), *Nucl. Fusion,* vol. 41, p. 197, 2001.

[7] H. Zohm; G. Gantenbein; A. Gude; S. Günter; F. Leuterer; M. Maraschek; J. Meskat; W. Suttrop; Q. Yu; ASDEX Upgrade Team; ECRH-Group (AUG), *Phys. Plasmas,* vol. 8, p. 2009, 2001.

[8] F. Leuterer, R. Dux, G. Gantenbein, J. Hobirk, A. Keller, K. Kirov, M. Maraschek, A. Mück, R. Neu, A.G. Peeters et al, *Nucl. Fusion,* vol. 43, p. 1329, 2003.

[9] R. J. La Haye, S. Günter, D. A. Humphreys, J. Lohr, T. C. Luce, M. E. Maraschek, C. C. Petty, R. Prater, J. T. Scoville and E. J. Strait, *Phys. Plasmas,* vol. 9, p. 2051, 2002.

[10] R. Prater, R. J. L. H. J. Loh, T. C. L. C. C. Petty, J. R. Ferron, D.A.Humphreys, E. J. Strait, F. W. Perkins and a. R. W. Harvey, *Nucl. Fusion,* vol. 43, p. 1128, 2003.

[11] A Isayama, Y Kamada, S Ide, K Hamamatsu, T Oikawa, T Suzuki, Y Neyatani, T Ozeki, Y Ikeda, K Kajiwara and the JT-60 team, *Plasma Phys. Controlled Fusion,* vol. 42, p. L37, 2000.

[12] A. Isayama, Y. Kamada, N. Hayashi, T. Suzuki, T. Oikawa, T. Fujita, T. Fukuda, S. Ide, H. Takenaga, K. Ushigusa et al, *Nucl. Fusion,* vol. 43, p. 1272, 2003.

[13] G. Ramponi, D. Farina, M. A. Henderson, F. Poli, G. Saibene and H. Zohm, *Fusion Sci. Technol.,* vol. 52, p. 193–201, 2007.

[14] L. Figini, D. Farina, M. Henderson, A. Mariani and G. S. E. Poli, *Plasma Phys. Controlled Fusion,* vol. 57, p. 054015, 2015.

[15] F. Poli, E. Fredrickson, M. Henderson, S-H. Kim, N. Bertelli, E. Poli, D. Farina and L. Figini, *Nucl. Fusion,* vol. 58, p. 016007, 2018.

[16] A. H. Reiman and N. J. Fisch, *Phys. Rev. Lett.,* vol. 121, p. 225001, 2018.

[17] D. De Lazzari and E. Westerhof, *Nucl. Fusion,* vol. 49, p. 075002, 2009.

[18] E. Rodriguez, A. H. Reiman and N. J. Fisch, *Phys. Plasmas,* vol. 26, p. 092511, 2019.

[19] E. Rodriguez, A. H. Reiman and N. J. Fisch, *Phys. Plasmas,* vol. 27, p. 042306, 2020.

[20] S. Jin, N. J. Fisch and A. H. Reiman, *Phys. Plasmas,* vol. 27, p. 062508, 2020.

[21] S. Frank, A. H. Reiman, N. J. Fisch and P. Bonoli, *Nucl. Fusion,* vol. 60, p. 096027, 2020.

[22] A. Reiman, N. Bertelli, N. Fisch, S. J. Frank; S. Jin, R. Nies and E. Rodriguez, *Phys. Plasmas,* vol. 28, p. 042508, 2021.

[23] S. Jin, A. H. Reiman and N. J. Fisch, *Phys. Plasmas,* vol. 28, p. 052503, 2021.

[24] R. Nies, A. H. Reiman, E. Rodriguez, N. Bertelli and N. J. Fisch, *Phys. Plasmas,* vol. 27, p. 092503, 2020.

[25] C. F. F. Karney and N. J. Fisch, *Nucl. Fusion,* vol. 21, p. 1549, 1981.

[26] C. F. F. Karney and N. J. Fisch, *Phys. Fluids,* vol. 28, p. 116, 1985.

[27] N. J. Fisch, *Rev. Mod. Phys.,* vol. 59, p. 175, 1987.

[28] C. F. F. Karney and N. J. Fisch, *Phys. Fluids,* Vols. 22,, p. 1817, 1979.

[29] C. F. F. Karney, N. J. Fisch and F. C. Jobes, *Phys. Rev. A,* vol. 32, p. 2554, 1985.

[30] I. Fidone, G. Granata and R. L. Meyer, *Phys. Fluids,* vol. 25, p. 2249, 1982.

[31] R. Fitzpatrick, *Phys. Plasmas,* vol. 2, p. 825, 1995.

[32] S. Inagaki, N. Tamura, K. Ida, Y. Nagayama, K. Kawahata, S. Sudo, T. Morisaki, K. Tanaka, T. Tokuzawa and the LHD Experimental Group, *Phys. Rev. Lett.,* vol. 92, p. 055002, 2004.

[33] G.W. Spakman, G.M.D. Hogeweij, R.J.E. Jaspers, F.C. Schuller, E. Westerhof, J.E. Boom, I.G.J. Classen, E. Delabie, C. Domier, A.J.H. Donne et al, *Nucl. Fusion,* vol. 48, p. 115005, 2008.

[34] K. Ida, K. Kamiya, A. Isayama and Y. Sakamoto, *Phys. Rev. Lett.,* vol. 109, p. 065001, 2012.

[35] L. Bardoczi, T. L. Rhodes, T. Carter, N. A. Crocker, W. A. Peebles and B. A. Grierson, *Phys. Plasmas,* vol. 23, p. 052507, 2016.

[36] A. W. Hornsby, M. Siccinio, A. G. Peeters, E. Poli, A. P. Snodin, F. J. Casson, C. Y. and G. Szepesi, *Plasma Phys. Controlled Fusion,* vol. 53, p. 054008, 2011.

[37] X Garbet, P Mantica, C Angioni, E Asp, Y Baranov, C Bourdelle, R Budny, F Crisanti, G Cordey, L Garzotti et al, *Plasma Phys. Control. Fusion,* vol. 46, p. B557, 2004.

[38] X Garbet, P Mantica, F Ryter, G Cordey, F Imbeaux, C Sozzi, A Manini, E Asp, V Parail, R Wolf and the JET EFDA Contributors, *Plasma Phys. Control Fusion,* vol. 46, p. 1351, 2004.

[39] E. Poli, Tardini, H. Zohm, E. Fable, D. Farina, L. Figini, N.B. Marushchenko and L. Porte, *Nucl. Fusion,* vol. 53, p. 013011, 2013.

[40] X. Chen, C. Petty, J. Lohr, D. Su, R. Prater, M. Cengher, M. Austin, C. Holcomb, L. Lao, R.I. Pinsker et al, *Nucl. Fusion,* vol. 62, p. 054001, 2022.

[41] Xi Chen, C.C. Petty, J. Lohr, R. Prater, M. Cengher, Y. Gorelov, L. Lao, D. Ponce, R.I. Pinsker, D. Su et al, in *28th IAEA Fusion Energy Conference*, 2021.